\documentstyle[preprint,aps,psfig]{revtex}
\begin{document}


\begin{center} 
DIFFERENTIAL EQUATIONS EXTENDED TO SUPERSPACE  
\end{center}


\begin{center}
J. Torres$^1$ and H.C. Rosu$^{1,2}$
\end{center}
\begin{center}
$^{1}$ Instituto de F\'{\i}sica, Universidad de Guanajuato,\\
Apdo Postal E-143, Le\'on, Guanajuato, M\'exico\\
$^{2}$ Instituto Potosino de Investigaci\'on Cient\'{\i}fica y Tecnol\'ogica,\\ 
Apdo Postal 3-74 Tangamanga, San Luis Potos\'{\i}, S.L.P., M\'exico
\end{center}

\bigskip
\bigskip

\noindent
We present a simple SUSY ${\mathcal N} _{S}$=2 superspace extension of the differential 
equations in which the sought solutions are considered to be 
real superfields but maintaining the common derivative operators and the 
coefficients of the differential equations unaltered. In this way, we
get selfconsistent systems of coupled differential equations for the 
components of the superfield. 
This procedure is applied to the Riccati equation, for which we obtain
in addition the system of coupled equations corresponding to the 
components of the general superfield solution.

\bigskip
\bigskip  

{\large \it Introduction}
\bigskip

\noindent
In the framework of quantum cosmology, Obregon et al
\cite{o} developed a formalism to find supersymmetric 
actions based on local ${\mathcal N}_{S}$=2 SUSY transformations and the 
concept of superfield. In particular, they generalized 
the local time transformations, $\delta t=a(t)$ that
leave invariant the 
action, to local grassmannian transformations involving
the Grassmann time variables $\eta$ and $\bar{\eta}$
$$
\delta t=a(t)+\frac{i}{2}\eta \beta (t)+ \frac{i}{2}\bar{\eta}\bar{\beta}
(t)
$$
$$
\delta \eta=\frac{1}{2}\bar{\beta}(t)+ \frac{1}{2}(\dot{a}(t)+ib(t))\eta+\frac{i}{2}\dot{{\bar{\beta}}}\eta \bar{\eta}~,
$$

$$
\delta \bar{\eta}=\frac{1}{2}\bar{\beta}(t)+ \frac{1}{2}(\dot{a}(t)-ib(t))\bar{\eta}-\frac{i}{2}\dot{\beta}
(t)\eta \bar{\eta}~,
$$
where $\eta$ is a grassmannian time coordinate and $\bar{\eta}$ is its
complex conjugate (we actually take $\eta\propto \theta _1 +i\theta _2$
and $\bar{\eta}\propto \theta _1 -i\theta _2$). The parameter $\beta(t)$
is the Grassmann complex parameter corresponding to the local SUSY ${\mathcal N}_{S}$=2  whereas $b(t)$ is the parameter of the local U(1) rotation group of $\eta$.
Because of these generalized transformations the ordinary fields turn into
superfields, i.e., $f(t)\rightarrow {\mathcal F}(t,\eta , \bar{\eta})$.
We shall consider only real superfields, ${\mathcal F}^{\dagger}={\mathcal F}$, where the dagger operation is defined
for grassmannian variables as $(\eta \bar{\eta})^{\dagger}=\bar{\eta}^{\dagger}\eta ^{\dagger}$.
Thus, for an arbitrary superfield
$$
{\mathcal N}(t,\eta,\bar{\eta})=N(t)+i\eta \bar{\psi}(t)+i\bar{\eta}\psi(t)+\eta \bar{\eta}V(t)
$$
one has
$$  
{\mathcal N}^{\dagger}(t,\eta,\bar{\eta})=N^{\dagger}(t)-i \bar{\psi}^{\dagger}(t)\eta ^{\dagger}-i\psi ^{\dagger}(t)\bar{\eta}^{\dagger}+ \bar{\eta}^{\dagger}\eta ^{\dagger}V^{\dagger}(t)~.
$$
The condition of reality implies
$$
N^{\dagger}=N ,\quad  \bar{\psi}^{\dagger}=\psi ,\quad V^{\dagger}=V,
\quad \eta ^{\dagger}=\bar{\eta}~.
$$

\bigskip

{\large \it Differential equations extended to superspace}

\bigskip

\noindent
There are many works on superspace extensions of differential
equations \cite{fluid}. We consider here one of the simplest possible extensions by turning the dependent variable of any differential
equation of arbitrary order $n$ into a superfield but without changing
the derivative operator as performed in the literature. Moreover,
we maintain unchanged the coefficients of the superextended equation because
in this way we get selfconsistent differential systems of coupled equations for the superfield components.   

Thus, what we are doing is to take the general equation
$$
a_n(t)y^{(n)}(t)+...+a_0(t)y(t)=F(t)
$$
and change it into
$$
a_n(t){\mathcal Y}^{(n)}(t, \eta , \bar{\eta})+...+a_0(t){\mathcal Y}(t, \eta , \bar{\eta})=F(t)~,
$$
where the superscript $(n)$ means the $n$th order derivative $\frac{d^n}{dt^n}$.

\bigskip

{\large \it Riccati equation}

\bigskip

\noindent
Our preferred example is the Riccati equation
$$
\frac{dy}{dt}=a(t)y^2+b(t)y+c(t)~.
$$
It is well known that the theory of Riccati equation is equivalent to
the theory of the linear differential equations of the second order.
This is due to the connection between the two through the direct transformation
$$
y(t)=-\frac{1}{a(t)}\frac{\dot{w}}{w}
$$
leading to
$$
\ddot{w}(t)-\left(\frac{\dot{a}}{a}+b(t)\right)\dot{w}(t)+a(t)c(t)w(t)=0
$$
and the inverse one
$$
w(t)=\exp\left(-\int a(t)y(t)dt \right)
$$
by which one goes back to the Riccati equation.
Another important result is the possibility to construct the general
Riccati solution once one knows a particular solution through the 
famous Bernoulli "ansatz"
$$
y_g=y_p+\frac{1}{f}~,
$$
which introduced in the Riccati equation turns it into the following linear 
equation 
$$
\dot{f}=-\Big[b(t)+2a(t)y_p(t)\Big]f-a(t)~.
$$
The solution of the latter is
$$
f(t)=\frac{1}{I(t)}\Big[-\int ^{t}a(x)I(x)dx+C\Big]~,
$$
where $I(t)=\exp\left(\int ^{t}(b(x)+2a(x)y_p(x))dx\right)$.

\bigskip

{\large \it Extended Riccati equation}

\bigskip

\noindent
We now write the superspace extended Riccati equation as follows 
\begin{equation} \label{ch1}
\frac{i}{2}\left\{D_{\eta},D_{\bar{\eta}}\right\}{\mathcal Y}(t,\eta,\bar{\eta})=a(t){\mathcal Y}^2(t,\eta,\bar{\eta})+b(t){\mathcal Y}(t,\eta,\bar{\eta})+c(t)
\end{equation}
where
$$
\frac{i}{2}\left\{D_{\eta},D_{\bar{\eta}}\right\}=\frac{d}{dt}
$$
and
$$
D_{\eta}=\frac{\partial}{\partial\eta}+i\bar{\eta}\frac{\partial}{\partial t} ,\quad \quad \quad D_{\bar{\eta}}=-\frac{\partial}{\partial\bar{\eta}}-i\eta \frac{\partial}{\partial t}~,
$$
with $D_{\eta}$ and $D_{\bar{\eta}}$ the supercovariant derivatives.\\
We write the superfield ${\mathcal Y}$ in the form 
\begin{equation} \label{ch2}
{\mathcal Y}(t,\eta,\bar{\eta})=y(t)+i\eta\bar{\lambda}(t)+i\bar{\eta}\lambda (t)+\eta\bar{\eta} G(t)
\end{equation}
Introducing (\ref{ch2}) in (\ref{ch1}) and identifying the corresponding
components one gets the following system
\begin{eqnarray}
\dot{y}&=&ay^2+by+c\\
\dot{\lambda}&=&(2ay+b)\lambda\\
\dot{\bar{\lambda}}&=&(2ay+b)\bar{\lambda}\\
\dot{G}&=&(ay+b)G+2a\bar{\lambda}\lambda
\end{eqnarray}
Since the two equations for the fermionic components are conjugate to one another it is
sufficient to solve only one of them.

In parallel with the ordinary procedure we propose the generalized 
transformation
\begin{equation}\label{ch3}
{\mathcal Y}=-\frac{1}{a}\frac{\dot{{\mathcal N}}}{{\mathcal N}}=-\frac{1}{a}{\mathcal N}^{-1}\dot{{\mathcal N}}
\end{equation}
where ${\mathcal N}^{-1}$ is defined by ${\mathcal N}^{-1}{\mathcal N}=
{\mathcal N}{\mathcal N}^{-1}=1$.
Considering ${\mathcal N}(t,\eta,\bar{\eta})=N(t)+i\eta \bar{\psi}(t)+i\bar{\eta}\psi(t)+\eta \bar{\eta}V(t)$ how can one obtain
the superfield ${\mathcal N}^{-1}$ ? To answer this question, we write
 ${\mathcal N}$ as ${\mathcal N}=N+\Lambda$, where $\Lambda=
i\eta \bar{\psi}(t)+i\bar{\eta}\psi(t)+\eta \bar{\eta}V(t)$.
Since ${\mathcal N}^{-1}=\frac{1}{{\mathcal N}}=\frac{1}{N+\Lambda}=
\frac{1}{N(1+\frac{\Lambda}{N})}$, expanding the latter expression we get
\begin{equation}\label{ch4}
\frac{1}{N(1+\frac{\Lambda}{N})}=\frac{1}{N}\sum_{k=0}^{\# {\scriptscriptstyle{\rm supercharges}}}
(-1)^k\left(\frac{\Lambda}{N}\right)^k=\frac{1}{N}\left(1-\frac{\Lambda}{N}
+\left(\frac{\Lambda}{N}\right)^2\right)~.
\end{equation}
Substituting $\Lambda$ in the latter formula one eventually gets
\begin{equation}\label{ch5}
{\mathcal N}^{-1}=\frac{1}{N}-i\eta N^{-2}\bar{\psi}-i\bar{\eta}N^{-2}\psi
+\eta \bar{\eta}(2N^{-3}\bar{\psi}\psi -N^{-2}V)~.
\end{equation}
Using (\ref{ch5}) and (\ref{ch3}) in (\ref{ch1}) one can get after some
tedious algebra the second order linear differential equations for all
the componets of the superfield ${\mathcal N}$. For the first bosonic
component we obtain naturally the standard equation
\begin{equation}\label{ch6}
\ddot{N}-\left(\frac{\dot{a}}{a}+b\right)\dot{N}+acN=0~.
\end{equation}
For the $\psi$ component we get
\begin{equation}\label{ch7}
\ddot{\psi}+\left(-a^{-1}\dot{a}-N^{-2}\dot{N}+N^{-1}\dot{N}-b\right)\dot{\psi}+\left(a^{-1}\dot{a}N^{-1}\dot{N}-N^{-1}\ddot{N}+b\right)
=0~,
\end{equation}
and identically for the $\bar{\psi}$ component in view of the reality
condition. On the other hand,
the equation for the $V$ component is more complicated and we do not 
include it here. We mention that it is possible to write the inverse 
generalized transformation to recover from these equations the extended 
Riccati one. One should write
\begin{equation}\label{ch8}
{\mathcal N}=e^{-\int ^{t}a{\mathcal Y}dt}
\end{equation}
and using ${\mathcal Y}=y+\Gamma (t, \eta , \bar{\eta})$ we have
\begin{equation}\label{ch9}
{\mathcal N}=e^{-\int ^{t}aydt}\Bigg[1-\int ^{t}a\Gamma dt+\frac{1}{2}
\int ^{t}\int ^{t}a(u)a(v)\Gamma(u,\eta , \bar{\eta})\Gamma(v,\eta , \bar{\eta})du dv\Bigg]~.
\end{equation}

\bigskip

{\large \it The general superfield solution of the extended Riccati equation}

\bigskip

\noindent
For the general superfield we proceed again by analogy with the 
standard calculus. We write
\begin{equation}\label{ch10}
{\mathcal Y}={\mathcal B}+{\mathcal D}^{-1}~,
\end{equation}
where ${\mathcal B}$ is a superfield satisfying the extended Riccati 
equation
\begin{equation} \label{ch11}
\frac{i}{2}\left\{D_{\eta},D_{\bar{\eta}}\right\}{\mathcal B}(t,\eta,\bar{\eta})=a(t){\mathcal B}^2(t,\eta,\bar{\eta})+b(t){\mathcal B}(t,\eta,\bar{\eta})+c(t)
\end{equation}
Which is the superequation ${\mathcal D}$ satisfies ? Substituting (\ref{ch10}) in
the extended Riccati equation we get
\begin{equation} \label{ch12}
\frac{i}{2}\left\{D_{\eta},D_{\bar{\eta}}\right\}
({\mathcal B}+{\mathcal D}^{-1})=a(t)({\mathcal B}^2+2{\mathcal B}{\mathcal D}^{-1}+{\mathcal D}^{-2})+b(t)({\mathcal B}+{\mathcal D}^{-1})+c(t)~,
\end{equation}
where we have used the fact that two superfields commute. Since $\mathcal{B}$ satisfies the Riccati superequation we can simplify
(\ref{ch12}) to the form
\begin{equation} \label{ch13}
-{\mathcal D}^{-2}\dot{{\mathcal D}}=2a{\mathcal B}{\mathcal D}^{-1}+b{\mathcal D}^{-1}+a{\mathcal D}^{-2}~.
\end{equation}
Multiplying by $-{\mathcal D}^{2}$ we obtain
\begin{equation} \label{ch14}
\dot{{\mathcal D}}=-(2a{\mathcal B}+b){\mathcal D}-a~,
\end{equation}
which is completely equivalent to the ordinary case.
We used $\frac{d}{dt}{\mathcal D}^{-1}=-{\mathcal D}^{-2}\dot{{\mathcal D}}$.

A case of special interest is $b=0$ that corresponds to 
the Riccati-Schroedinger connection in supersymmetric quantum mechanics. For this case we get the system
\begin{eqnarray}
\dot{B}&=& aB^2+c\\
\dot{\varphi}&=&2aB\varphi\\
\dot{\bar{\varphi}}&=&2aB\bar{\varphi}\\
\dot{A}&=&2aBA+2a\bar{\varphi}\varphi \\
\bigskip
\dot{D}&=&-2aBD-a\\
\dot{\psi}&=&-2a(B\psi +D\varphi)\\
\dot{\bar{\psi}}&=&(B\bar{\psi} +D\bar{\varphi})\\
\dot{U}&=&-2a(BU+\bar{\varphi}\psi +\bar{\psi}\varphi +AD)
\end{eqnarray}
where 
$$
{\mathcal B}(t,\eta,\bar{\eta})=B+i\eta\bar{\varphi}+i\bar{\eta}\varphi+\eta\bar{\eta} A ,
$$ 
$$
{\mathcal D}(t,\eta,\bar{\eta})=D+i\eta\bar{\psi}+i\bar{\eta}\psi +\eta\bar{\eta} U .
$$

\bigskip

{\large \it Conclusion}

\bigskip

\noindent
Although the procedure of superspace extension presented here is quite 
simple, it appears to be quite interesting for physical applications
related to supersymmetric quantum mechanics and other fields of 
physics. Recently, we have used this scheme for the Riccati equation of constant coefficients
that appears in FRW barotropic cosmologies for which we have solved
explicitly the system of coupled differential equations generated by
our procedure \cite{rt}. In addition, we have interpreted the obtained solutions
as the components of a Hubble superfield parameter.

\end{document}